\documentclass[10pt]{iopart}
\usepackage{latexsym}
\usepackage{amsfonts}
\usepackage{amsbsy}
\usepackage{mathrsfs}

\begin{document}

\title[The boundary field theory induced by the
Chern-Simons theory]{The boundary field theory induced by the Chern-Simons
theory}

\author{Alejandro Gallardo and Merced Montesinos}

\address{Departamento de F\'{\i}sica, Cinvestav, Instituto Polit\'ecnico Nacional 2508,
San Pedro Zacatenco, 07360, Gustavo A. Madero, Ciudad de M\'exico, M\'exico.}
\eads{\mailto{agallardo@fis.cinvestav.mx}, \mailto{merced@fis.cinvestav.mx}}

\begin{abstract}
The Chern-Simons theory defined on a 3-dimensional manifold with boundary is
written as a two-dimensional field theory defined only on the boundary of the
three-manifold. The resulting theory is, essentially, the pull-back to the
boundary of a symplectic structure defined on the space of auxiliary fields in
terms of which the connection one-form of the Chern-Simons theory is expressed
when solving the condition of vanishing curvature. The counting of the
physical degrees of freedom living in the boundary associated to the model is
performed using Dirac's canonical analysis for the particular case of the
gauge group $SU(2)$. The result is that the specific model has one physical
local degree of freedom. Moreover, the role of the boundary conditions on the original
Chern-Simons theory is displayed and clarified in an example, which shows how the
gauge content as well as the structure of the constraints of the induced boundary theory is
affected.
\end{abstract}


\submitto{\JPA}

\maketitle

\section{Introduction}
As is well-known, Chern-Simons theory defined on a three-dimensional manifold
${\mathscr{M}}^3$ {\it without} boundary has no local degrees of freedom
\cite{naka,blago,carlip}. The only physical degrees of freedom of the theory are global and
associated with the topological properties of the three-manifold. Nevertheless,
when the three-manifold ${\mathscr{M}}^3$ has a boundary $\partial
{\mathscr{M}}^3$ then the theory has local physical degrees of freedom living
at $\partial {\mathscr{M}}^3$ . This fact is also very well-known
in the literature \cite{carlip,moore,banados}. In spite of this, the origin and
the meaning of the physical degrees of freedom has not been completely elucidated,
as far as we know.

The purpose of this paper is to make a contribution to this end.
Our analysis does not hold for arbitrary three-dimensional manifolds with
boundary. However, it has the advantage of displaying without ambiguities
the field theory living at the boundary when the current analysis holds. More
precisely, the goal of this paper is to construct a {\it covariant} action
principle living at the boundary $\partial {\mathscr{M}}^3$ of the
three-manifold ${\mathscr{M}}^3$ by solving the equations of motion $F=0$
and expressing the connection field in terms of auxiliary fields. The resulting
action principle for the field theory living at $\partial {\mathscr{M}}^3$  will
depend functionally on these auxiliary fields. Once we have at our hand the
covariant action at $\partial {\mathscr{M}}^3$, we count the number of local
degrees of freedom of the theory by employing Dirac's canonical analysis.

It is worth noticing that the usual approaches for studying the local degrees
of freedom at the boundary involve the following features \cite{moore,banados,eli,cou}:
(1) they explicitly break covariance of the connection one-form, (2) related
with item (1) is the fact that solutions to Dirac constraints are used to
build the boundary action, (3) the handle of boundary terms and boundary
conditions is not systematic. On the contrary, our approach leads to the
construction of a covariant action principle. In our opinion, the knowledge
of the Lagrangian action principle is relevant for various reasons, among them:
it can be used to couple other interactions to it or to express already known
actions (for instance, Palatini action in three-dimensional general
relativity) in terms of it. Moreover, in our method, boundary terms and
boundary conditions can be systematically implement through Dirac's canonical
analysis as primary constraints [see, however, \cite{freed,freed2,sengupta}].

\section{The boundary field theory induced by an $SU(2)$ Chern-Simons theory}
The starting point is the action principle
\begin{eqnarray}\label{cs}
S[A^i] = \frac{\kappa}{4\pi} \int_{\mathscr{M}^3} \left [ A^i \wedge F^j -
\frac{1}{3!} f^i\,_{kl} A^k \wedge A^l \wedge A^j \right ] k_{ij},
\end{eqnarray}
where $F^i = d A^i + \frac12 f^i\,_{jk} A^j \wedge A^k$ is the curvature of
the connection one-form $A= A^i J_i$, $J_i$ are the generators of the Lie
algebra and satisfy $[J_i, J_j]= f^k\,_{ij} J_k$ with $f^k\,_{ij}$ the
structures constants of the Lie algebra, $k_{ij}= -\frac12 f^k\,_{il}
f^l\,_{jk}$ is the Killing-Cartan metric and the constant $\kappa$ is the so-called
level of the theory. It is assumed that $\mathscr{M}^3$
has a boundary, $\partial {\mathscr{M}}^3 \neq \emptyset$.

The variation of the action principle (\ref{cs}) respect to the gauge
connection $A$ gives rise to the equations of motion $F^i=0$. This tells us that
the space of solutions is the space of flat connections. Consequently, the theory
has no local physical degrees of freedom in the bulk and any possible local
physical degree of freedom is contained at the boundary \cite{blago,carlip,moore}.

For the sake of simplicity, let us restrict the analysis to the gauge group
$SU(2)$. In particular, the Lie algebra is spanned by the traceless
skew-Hermitian $2\times 2$ matrices $J_i= -\frac{i}{2} \sigma_i$, $i=1,2,3$,
which satisfy $[J_i , J_j]= \varepsilon^k\,_{ij} J_k$ where
$\varepsilon^k\,_{ij}$ are the structure constants
($\varepsilon^1\,_{23}=-\varepsilon^2\,_{13}=\varepsilon^3\,_{12}=1$) and
$\sigma_i$ are the Pauli matrices; $su(2)$ indices $i,j,k,\ldots$ are raised
and lowered with the Killing-Cartan metric $\delta_{ij} = -\frac12
\varepsilon^k\,_{il} \varepsilon^l\,_{jk}$, $\mbox{diag}
(\delta_{ij})=(1,1,1)$. Under these assumptions it follows that
\begin{eqnarray}
U = e^{X^i J_i} = I \cos{\mid \mid X \mid\mid} +
2 {\hat X}^i J_i \sin{\mid \mid X\mid\mid},
\end{eqnarray}
is an element of $SU(2)$ where $\mid \mid X \mid\mid = \sqrt{X^i X^j
\delta_{ij}}$, ${\hat X}^i = \frac{X^i}{\mid\mid X\mid\mid}$, and $I$ is the
$2 \times 2$ identity matrix. Using this to compute the $su(2)$ connection
1-form $\bar{A} \equiv {\bar A}^i J_i = U^{-1} dU$, we get
\begin{eqnarray}\label{solution}
\fl
{\bar A}^i = \frac{\sin{||X||}}{||X||} d X^i +
\frac{\left ( ||X|| - \sin{||X||} \right )}{||X||^2} X^i d ||X|| -
\frac{2}{||X||^2} \sin^2{\left ( \frac{||X||}{2} \right )}
\varepsilon^i\,_{jk} X^j d X^k.
\end{eqnarray}
By construction it follows that the curvature of ${\bar A}^i$ vanishes, ${\bar
F}^i = d {\bar A}^i + \frac12 \varepsilon^i\,_{jk} {\bar A}^j \wedge {\bar
A}^k=0$. Therefore, on-shell (i.e., using the equations of motion ${\bar
F}^i=0$) the Chern-Simons action principle (\ref{cs}) acquires the form
\begin{eqnarray}
S[{\bar A}] = -\frac{\kappa}{4\pi} \frac{1}{3!} \int_{{\mathscr{M}}^3} \varepsilon_{ijk} {\bar A}^i
\wedge {\bar A}^j \wedge {\bar A}^k,
\end{eqnarray}
which can be rewritten using (\ref{solution}) as
\begin{eqnarray}\label{key}
S &=& \frac{\kappa}{2 \pi} \frac{1}{3!} \int_{{\mathscr{M}}^3}
\frac{\left ( \cos{||X||}-1 \right )}{||X||^2}
\varepsilon_{ijk} d X^i \wedge d X^j \wedge d X^k \nonumber\\
&=& \frac{\kappa}{4 \pi} \int_{{\mathscr{M}}^3} d \left [ \frac{\left ( \sin{||X||}- ||X|| \right )}{||X||^3}
\varepsilon_{ijk} X^i d X^j \wedge d X^k + d H \right ] \nonumber\\
& =& \frac{\kappa}{4\pi} \int_{\partial {\mathscr{M}}^3} \left [ \frac{\left ( \sin{||X||}- ||X|| \right )}{||X||^3}
\varepsilon_{ijk} X^i d X^j \wedge d X^k + d H \right ],
\end{eqnarray}
and because $\partial {\mathscr{M}}^3$ has no boundary, $\partial \left
(\partial {\mathscr{M}}^3 \right )=\emptyset$, we finally get
\begin{eqnarray}\label{nlsm}
S[X^i]&=& \frac{\kappa}{4\pi} \int_{\partial {\mathscr{M}}^3}
\frac{\left ( \sin{||X||}- ||X|| \right )}{||X||^3}
\varepsilon_{ijk} X^i d X^j \wedge d X^k.
\end{eqnarray}
This is the action principle induced by the Chern-Simons theory when the
latter is defined on a three-manifold $\mathscr{M}^3$ with a boundary
$\partial {\mathscr{M}}^3$. In order to get the second line of Eq. (\ref{key})
we have assumed that the third cohomology group $H^3$ of ${\mathscr{M}}^3$
vanishes, $H^3 \left ( {\mathscr{M}}^3 \right ) = \emptyset$, which guarantees
that the three-form in the integrand of the first line of Eq. (\ref{key}) is
globally exact \cite{naka}.

Before going on, notice that the model (\ref{nlsm}) belongs to the class of
theories defined by the action principle
\begin{eqnarray}\label{nlsm+}
S[X^i]&=& \frac{\kappa}{4 \pi} \int_{\partial {\mathscr{M}}^3} f(||X||)\,
\varepsilon_{ijk} X^i d X^j \wedge d X^k,
\end{eqnarray}
for the particular choice of $f$ given by
\begin{eqnarray}
f(||X||) = \frac{\sin{||X||}- ||X||}{||X||^3}.
\end{eqnarray}
The model is, essentially, the integral of the pull-back of the symplectic
structure $\omega = f(||X||)\, \varepsilon_{ijk} X^i d X^j \wedge d X^k$ to
the boundary $\partial {\mathscr{M}}^3$. The symplectic structure $\omega$ can
be thought as living in $\mathbb{R}^3$, which is isomorphic to $su(2)$-the Lie
algebra of $SU(2)$-and whose points are labeled by the coordinates $X^i$. The
symplectic structure $\omega$ is degenerate because it is a $3\times3$
antisymmetric matrix.

In order to count the physical degrees of freedom of the field theory defined
by the action principle (\ref{nlsm}), Dirac's canonical analysis is employed.
It is convenient to perform the analysis using (\ref{nlsm+}) instead of
(\ref{nlsm}). We take ${\mathscr{M}}^3= \mathbb{R} \times D^2 $ where $D^2$ is
a two-dimensional disc. Therefore, $\partial {\mathscr{M}}^3 = \mathbb{R}
\times S^1$ where the circle $S^1$ is the boundary of $D^2$, $S^1= \partial
D^2$. Note that $H^3 \left (\mathbb{R} \times D^2 \right ) =
\emptyset $ \cite{naka}. Let ($\tau$, $\sigma$) be local coordinates that label the points of
$\mathbb{R} \times S^1$, the time coordinate $\tau$ labels the points of
$\mathbb{R}$ and the space coordinate $\sigma$ labels the points of $S^1$.
Therefore, $d X^i =
\partial_{\tau} X^i d \tau + \partial_{\sigma} X^i d \sigma \equiv {\dot
X}^i d \tau +
\partial_{\sigma} X^i d \sigma$ and the action (\ref{nlsm+}) becomes
\begin{eqnarray}\label{wzw action}
S[X^i] = \frac{\kappa}{2 \pi} \int_{\mathbb{R}} d \tau \int_{S^1} d \sigma \, f(||X||)\,
\varepsilon_{ijk} X^i {\dot X}^j \partial_{\sigma} X^k.
\end{eqnarray}
The definition of the momenta $p_i$, canonically conjugate to $X^i$, implies
that we have three primary constraints
\begin{eqnarray}\label{cons}
\gamma_i &:=& p_i + \frac{\kappa}{2 \pi} f(||X||)\, \varepsilon_{ijk} X^j
\partial_{\sigma} X^k \approx 0.
\end{eqnarray}
A straightforward computation implies that the canonical Hamiltonian vanishes,
and so the action principle acquires the form
\begin{eqnarray}
S[X^i, p_i, \lambda^i] =  \int_{\mathbb{R}} d \tau \int_{S^1} d \sigma \,
\left ( {\dot X}^i p_i - \lambda^i \gamma_i \right ).
\end{eqnarray}
Computing the variation of this action with respect to the independent
variables we get the dynamical equations
\begin{eqnarray}\label{dyn}
{\dot X}^i &=& \lambda^i, \nonumber\\
{\dot p}_i &=& \frac{\kappa}{2 \pi} f \varepsilon_{ijk} \left ( 2 \lambda^j \partial_{\sigma} X^k
- X^j \partial_{\sigma} \lambda^k \right ) +
\frac{f'}{||X|| f} \lambda^j \left ( p_j - \gamma_j \right ) X_i  \nonumber \\
& & + \frac{\kappa}{2\pi} \left ( \partial_{\sigma} f \right ) \varepsilon_{ijk} \lambda^j X^k,
\end{eqnarray}
as well as the constraints (\ref{cons}). Here $f'$ stands for the derivative
of $f$ with respect to $||X||$.

Using Eqs. (\ref{dyn}) to compute the evolution of the primary constraints
$\gamma_i$, we obtain
\begin{eqnarray}
{\dot \gamma}_i &=& \frac{\kappa}{2\pi} \varepsilon_{ijk} \lambda^j
\left [ 3 f \partial_{\sigma} X^k + \left ( \partial_{\sigma} f \right ) X^k \right ] \nonumber\\
&& + \frac{f'}{||X|| f } \left [ \lambda^j (p_j - \gamma_j) X_i +
\lambda^j X_j \left( p_i - \gamma_i \right ) \right ],
\end{eqnarray}
and therefore the consistency conditions for the primary constraints ${\dot
\gamma}_i \approx 0$ imply a system of three linear and homogeneous equations
for the Lagrange multipliers $\lambda^i$, given by
\begin{eqnarray}\label{linear}
\frac{\kappa}{2 \pi} \varepsilon_{ijk} \lambda^j
\left ( 3f \partial_{\sigma} X^k + X^k \partial_{\sigma} f \right ) +
\frac{f'}{||X|| f} \left ( p_j X_i - p_i X_j \right ) \lambda^j  \approx 0.
\end{eqnarray}
This system of equations has one non-trivial null vector in the generic case,
whose components are given by
\begin{eqnarray}
v^i (\tau,\sigma) &=&\frac{1}{2} \int_{S^1} d \sigma' \varepsilon^{ijk}
\{ \gamma_j (\sigma) , \gamma_k (\sigma') \}, \nonumber\\
&=& \frac{\kappa}{2 \pi} \left ( 3 f \partial_{\sigma} X^i +
X^i \partial_{\sigma} f \right ) + \frac{f'}{||X|| f} \varepsilon^i\,_{jk} X^j p^k,
\end{eqnarray}
which gives rise to the first-class constraint
\begin{eqnarray}\label{difeos}
\gamma := v^i \gamma_i = \frac{\kappa}{2\pi}
\left ( 3f +  ||X|| f' \right ) p_i \partial_{\sigma} X^i \approx 0.
\end{eqnarray}
The meaning of the gauge symmetry generated by (\ref{difeos}) is clear because the
constraint (\ref{difeos}) is $\left ( 3f +  ||X|| f' \right )$ times the
diffeomorphism constraint. Thus, (\ref{difeos}) reflects the fact that the action for
the two-dimensional theory at the boundary (\ref{nlsm}) is
diffeomorphism invariant. As a consequence the rank of the system of
equations (\ref{linear}) is two, which means that there are two second-class
constraints among the $\gamma_i$. Thus, the counting of the physical degrees
of freedom is as follows. The extended phase space is parametrized by 3
configuration variables $X^i$ and the corresponding 3 canonical momenta
$\pi_i$, there are 1 first-class and 2 second-class constraints. Therefore
the system has $\frac12 (2\times 3 - 2 \times 1 -2 )=1$ physical degree of
freedom per point of $S^1$.

In summary, the physics involved in a Chern-Simons theory defined on a
three-manifold with boundary is completely different from the physics involved
in the Chern-Simons theory defined on a three-manifold without boundary. In
the former there are local and physical excitations living at the boundary
while in the latter there are simply no local excitations. It is just the
presence of the boundary that makes one theory be completely different from
the other, i.e., it is the boundary what generates-in the sense explained
above-the local dynamics at the boundary itself.

\section{Adding boundary conditions on the original connection}
Now, the issue of the boundary conditions will be analyzed. As we already mentioned, the
variation of the action principle (\ref{cs}) with respect to the connection
gives the equations of motion $F=0$ provided that
\begin{eqnarray}\label{frontera}
\int_{\partial {\mathscr{M}}^3} k_{ij}A^i\wedge \delta A^j,
\end{eqnarray}
vanishes. On the other hand, when the solution for the connection one-form
(\ref{solution}) for the particular case of $SU(2)$ is inserted back into the
action, the boundary action principle (\ref{nlsm}) is obtained. Even though
the original equations of motion $F=0$ are obtained keeping some specific boundary
conditions on $A^i$ such that (\ref{frontera}) vanishes, this information is
not incorporated into the resulting action principle (\ref{nlsm}). This
situation is similar to what happens, for instance, in the first order
formulation for three-dimensional gravity. There, the action principle
depends functionally on the triad $e^I$ and the Lorentz connection $\omega^I\,_J$
and it is given by
\begin{eqnarray}\label{palatini}
S[e^I, \omega^I\,_J] = \int_{{\mathscr{M}}^3} \varepsilon_{IJK} e^I \wedge R^{JK} [\omega],
\end{eqnarray}
where $R^I\,_J [\omega] = d \omega^I\,_J + \omega^I\,_K \wedge \omega^K\,_J$
is the curvature of the Lorentz connection $\omega^I\,_J$. The variation of the
action (\ref{palatini}) with respect to the independent fields gives the equations of motion
$d e^I + \omega^I\,_J \wedge e^J=0$ and $R^{IJ}[\omega]=0$ provided that
$\int_{\partial {\mathscr{M}}^3} \varepsilon_{IJK} e^I \wedge \delta \omega^{JK}$ vanishes.
When we solve $d e^I + \omega^I\,_J \wedge e^J=0$ for the connection $\omega^I\,_J$ in terms of the
triad, namely, $\omega^I\,_J [e]$ and plug it back into the action (\ref{palatini}), we
recover, essentially, the Einstein-Hilbert action $S[e]$ expressed in terms of the triads. There
is no need of adding to $S[e]$ the specific boundary conditions on the fields
of the original action principle (\ref{palatini}) that led to the original equations of motion.

Let us come back to the Chern-Simons theory. One might be interested in
keeping the original boundary conditions on the connection that kill
(\ref{frontera}) in the resulting field theory at the boundary
(\ref{nlsm}). Due to the fact the boundary action principle depends on the
auxiliary fields $X^i$, the original conditions on the connection must be
rewritten, using (\ref{solution}), as conditions on the auxiliary fields
$X^i$. By doing this, these boundary conditions are added to (\ref{nlsm}) via Lagrange
multipliers or as primary constraints if we are interested in the Hamiltonian
formulation of the theory. Let us apply these ideas in what follows:

As an illustration, we will take the boundary conditions
$A^i_{\tau} +\varepsilon A^i_{\sigma}=0$ on $\mathbb{R} \times S^1$ where
$\varepsilon=\pm1$, which kill (\ref{frontera}). We add these boundary conditions via Lagrange
multipliers $u^i$ to the action principle (\ref{nlsm+}), which becomes
\begin{eqnarray}
\fl S[X^i,u^i]&:=&\frac{\kappa}{2\pi}\int_{\mathbb{R}}d \tau \int_{S^1}
d\sigma\bigg[f\varepsilon_{ijk}X^i\dot{X}^j
\partial_\sigma X^k-\dot{X}^iu^j\bigg((3f+\|X\|f')\varepsilon_{ijk}X^k\nonumber\\
\fl&&+(\|X\|^2f+1)\delta_{ij}-fX_i X_j\bigg)-\varepsilon u^j\partial_\sigma X^i\bigg((3f+\|X\|f')
\varepsilon_{ijk}X^k\nonumber\\
\fl&&+(\|X\|^2f+1)\delta_{ij}-fX_iX_j\bigg)\bigg],
\end{eqnarray}
after making the $1+1$ decomposition. The definition of the momenta $(p_i,\pi_i)$,
canonically conjugated to the coordinates $(X^i,u^i)$, generates now six primary constraints,
\begin{eqnarray}\label{constraints-2}
 \phi_i&:=&p_i+\frac{\kappa}{2\pi}f\epsilon_{ijk}X^j
\partial_\sigma X^k+\frac{\kappa}{2\pi}u^j\bigg((3f+\|X\|f')\varepsilon_{ijk}X^k\nonumber\\
&&+ (\|X\|^2f+1)\delta_{ij}-fX_iX_j\bigg)\approx0,\\
 \chi_i&:=&\pi_i\approx0.\nonumber
\end{eqnarray}
A straightforward computation implies that the canonical Hamiltonian is
\begin{eqnarray}
H_0:=-\varepsilon p_i\partial_\sigma X^i,
\end{eqnarray}
and so the action principle acquires the form
\begin{eqnarray}
\fl S[X^i,u^i,p_i,\pi_i,\lambda^i,\Lambda^i]:=\int_{\mathbb{R}}
d\tau\int_{S^1}d\sigma\bigg(\dot{X}^ip_i+\dot{u}^i\pi_i+\varepsilon p_i
\partial_\sigma X^i-\lambda^i\phi_i-\Lambda^i\chi_i\bigg).
\end{eqnarray}
Its variation with respect to the independent variables yields the dynamical equations
\begin{eqnarray}\label{eq-mot}
\fl\dot{X}^i &=&\lambda^i-\varepsilon\partial_\sigma X^i, \nonumber\\
\fl \dot{u}^i &=& \Lambda^i,\nonumber\\
\fl\dot{\pi}_i &=&-\frac{\kappa}{2\pi}\lambda^j\bigg((3f+\|X\|f')\varepsilon_{jik}X^k+
(\|X\|^2f+1)\delta_{ij}-fX_iX_j\bigg), \nonumber\\
\fl\dot{p}_i&=&\frac{\kappa
}{2\pi}f\varepsilon_{ijk}\left(2\lambda^j\partial_\sigma X^k-X^j\partial_\sigma\lambda^k\right)
+\frac{\kappa}{2\pi}(\partial_\sigma f)\varepsilon_{ijk}\lambda^jX^k-
\frac{\kappa}{2\pi}\frac{f'}{\|X\|f}X^i\lambda^j(\phi_j-p_j)\nonumber\\
\fl &&-\varepsilon\partial_\sigma p_i-\frac{\kappa}{2\pi}\lambda^ju^k
\frac{\partial}{\partial X^i}\bigg((3f+\|X\|f')
\varepsilon_{jkl}X^l+(\|X\|^2f+1)\delta_{jk}-fX_jX_k\bigg)\nonumber\\
\fl &=&\frac{\kappa
}{2\pi}f\varepsilon_{ijk}\left(2\lambda^j\partial_\sigma X^k-X^j\partial_\sigma\lambda^k\right)
+\frac{\kappa}{2\pi}(\partial_\sigma f)\varepsilon_{ijk}\lambda^jX^k-
\frac{\kappa}{2\pi}\frac{f'}{\|X\|f}X^i\lambda^j(\phi_j-p_j)\nonumber\\
\fl &&-\varepsilon\partial_\sigma p_i-\frac{\kappa}{2\pi}\lambda^ju^k\bigg(
\left(\frac{4f'+\|X\|f''}{\|X\|}\right)\varepsilon_{jkl}X_iX^l+
\left(2f+\|X\|f'\right)X_i\delta_{jk}\nonumber\\
\fl &&+(3f+\|X\|f')\varepsilon_{ijk}-\frac{f'}{\|X\|}X_iX_jX_k-f(\delta_{ij}X_k+X_j\delta_{ik})\bigg),
\end{eqnarray}
as well as the constraints (\ref{constraints-2}).
By using the equations of motion (\ref{eq-mot}), the primary constraints
are evolved in time, which yields
\begin{eqnarray}\label{ultima}
\fl\dot{\phi}_i&=&\frac{\kappa}{2\pi}\varepsilon_{ijk}\lambda^j(3f\partial_\sigma X^k
+(\partial_\sigma f)X^k)+\frac{f'}{\|X\|f}\lambda^j(X_i(p_j-\phi_j)-X_j(p_i-\phi_i))\nonumber\\
\fl &&+\frac{\kappa}{2\pi}\lambda^ju^k\bigg(\left(\frac{(\|X\|f'-f)f'-
\|X\|ff''}{\|X\|f}\right)(\varepsilon_{jkl}X_i-\varepsilon_{ikl}X_j)X^l\nonumber\\
\fl &&-2(3f+\|X\| f')\varepsilon_{ijk}X^k-(\|X\|f'+(3+\|X\|^2)f+1)(X_i\delta_{jk}-X_j\delta_{ik})\bigg)
\nonumber\\
\fl&&+\frac{\kappa}{2\pi}\Lambda^j\bigg((3f+\|X\|f')\varepsilon_{ijk}X^k+
(\|X\|^2f+1)\delta_{ij}-fX_iX_j\bigg)-\varepsilon\frac{\kappa}{2\pi}\partial_\sigma(f\varepsilon_{ijk}X^j
\partial_\sigma X^k)\nonumber\\
\fl &&-\varepsilon\partial_\sigma p_i-\varepsilon\frac{\kappa}{2\pi}u^j\partial_\sigma
\bigg((3f+\|X\|f')\epsilon_{ijk}X^k+(\|X\|^2f+1)\delta_{ij}-fX_iX_j\bigg),\nonumber\\
\fl\dot{\chi}_i&=&-\frac{\kappa}{2\pi}\lambda^{j}\bigg((3f+\|X\|f')\epsilon_{jik}X^k
+(\|X\|^2f+1)\delta_{ji}-fX_jX_i\bigg),
\end{eqnarray}
and therefore by the consistency condition for primary constraints,
$\dot{\phi}\approx0$ and $\dot{\chi}\approx0$, we get a system of
six inhomogeneous linear equations for the Lagrange multipliers. A
straightforward computation shows that the determinant of the
corresponding matrix is equal to $\left ( \frac{k}{2\pi}\right)^6
\left [ (1+\|X\|^2f)^2+\|X\|^2(3f+\|X\|f')^2 \right ]^2$, which does
not vanish generically. Therefore, the six Lagrange multipliers can
be fixed: $\lambda^i=0$ and $\Lambda^i$ gets a cumbersome
expression, which is not displayed here but that involves the
inhomogeneous part of the system (\ref{ultima}). From this, we
conclude that there are no more constraints and that
(\ref{constraints-2}) are all second-class. Finally, the counting of
local degrees of freedom is as follows. Due to the fact that there
are six coordinates $(X^i,u^i)$ and their respective six momenta
($p_i,\pi_i$), then the system has $\frac{1}{2}\left [ 2 \times
(3+3) - 6 \right ] =3$ physical degrees of freedom per point on
$S^1$.

In summary, the example illustrates the role of boundary conditions on
the original boundary field theory (\ref{nlsm+}), namely, the specific choice of
the boundary conditions $A_\tau+\varepsilon A_\sigma =0$ increases the number of
local degrees of freedom from one to three. Moreover, these particular boundary
conditions kill the gauge freedom of the original boundary field theory. Therefore, we
conclude that depending on the specific form for the boundary conditions that we might impose
on the original connections, the number of physical degrees could be modified as well as the
gauge content of the original boundary theory.

\section{Implications: a method to construct field theories at the boundary}
One of the main lessons learned from the Chern-Simons theory defined on a
three-manifold with boundary is that the results of the previous section
provide a method to construct field theories living at the boundary. The
method can be summarized in the following steps:
\begin{enumerate}
\item Start with an action principle for a given theory defined on a
    $n$-dimensional manifold $\mathscr{M}^n$ with boundary $\partial
{\mathscr{M}}^n$.
\item Solve the equations of motion and write the Lagrangian $\bold{L}$ as
    $\bold{L}=d \bold{L}_{n-1}$. It is essential that the nth cohomology
    group vanishes in order for $\bold{L}= d \bold{L}_{n-1}$ be globally
    defined on $\mathscr{M}^n$.
\item Use Stokes theorem to build a theory defined on $\partial
    {\mathscr{M}}^{n-1}$ and given by $S=\int_{\partial \mathscr{M}^{n-1}}
    \bold{L}_{n-1}$.
\item Apply Dirac's canonical analysis to $S=\int_{\partial
    \mathscr{M}^{n-1}} \bold{L}_{n-1}$ in order to unreveal its physical
    local degrees of freedom.
\end{enumerate}
Whether the steps from (i) to (ii) can be carried out or not will depend on
the specific Lagrangian $\bold{L}$ under consideration, of course. We do not
have, at this moment, a criterium that allows us to say which Lagrangians
$\bold{L}$ will fulfil the properties (i) and (ii). However, if these steps
could be implemented in a given theory, this will allows us to study de direct
relation between the dynamic on the bulk and its correspondent dynamics on the
boundary in an unambiguous way.

\section{Concluding Remarks}
In this paper we have studied the boundary field theory induced
by the $SU(2)$ Chern-Simons theory when this is defined on a three-dimensional
manifold with boundary. The canonical analysis applied to the resulting action
principle for the theory at the boundary shows that it has one physical local degree of
freedom, whose origin is attached to the fact that the three-dimensional
manifold has a boundary; there are no (physical) local degrees of freedom
when the three-manifold has no boundary. Furthermore, the number of local degrees of freedom as
well as the gauge content of the theory can be changed by adding conditions on the
gauge potentials of the initial Chern-Simons theory to the action principle for
the boundary theory, which has been illustrated for a particular choice of
the boundary conditions. This last point is, certainly, not surprising but makes clear
the role of boundary conditions on the boundary field theory and the origin of the
boundary degrees of freedom of the Chern-Simons theory. Moreover, the techniques
used here can be applied to other Lie groups, besides $SU(2)$, and the systematic
procedure extracted from the model and described in section 4 can, in principle, be
applied to to other theories defined in arbitrary spacetimes with boundaries.

For instance, these ideas can be applied to Euclidean gravity in $(2+1)$ dimensions
with a negative cosmological constant $\Lambda$. The action principle for this theory
in the first order formalism
\begin{eqnarray}\label{palatinicc}
\fl S_{\Lambda}=\alpha\int_{\mathscr{M}^3}\left[e_I\wedge\left(d\omega^I
+\frac{1}{2}\varepsilon^I\,_{JK}\omega^J\wedge \omega^K\right)-\frac{\Lambda}{6}
\varepsilon_{IJK}e^I\wedge e^J\wedge e^K\right],
\end{eqnarray}
where $\alpha$ is a constant to adjust the units of the fields, can be expressed as
two ``independent'' Chern-Simons theories
\cite{carlip,arcioni,achucaro,witten} by first defining two connection one-forms
$A^I_{\pm}$ in terms of the triad $e^I$ and the Lorentz connection $\omega^I$ as follows
\begin{eqnarray}
A^I_\pm=\omega^I\pm\frac{1}{\ell} e^I,
\end{eqnarray}
with $\Lambda=-1/{\ell^2}$ and then plugging back $e^I$ and $\omega^I$ (in terms of $A^I\,_{\pm}$)
into the action principle (\ref{palatinicc}) to rewrite it as
\begin{eqnarray}\label{cs-g}
\fl S_{G\Lambda} [A_{+}, A_{-}] &=& \frac{\kappa}{4\pi}\int_{\mathscr{M}^3}\left[A^I_+\wedge F^J_+ -
\frac{1}{3!}f^I\,_{KL}A^K_+\wedge A^L_+\wedge A^J_+\right]k_{IJ} \nonumber\\
\fl &&- \frac{\kappa}{4\pi}\int_{\mathscr{M}^3}\left[A^I_-\wedge F^J_--
\frac{1}{3!}f^I\,_{KL}A^K_- \wedge A^L_-\wedge A^J_-\right]k_{IJ} \nonumber\\
\fl && - \frac{\kappa}{4\pi}\int_{\partial\mathscr{M}^3}A^I_+\wedge A^J_-k_{IJ},
\end{eqnarray}
$\kappa= \alpha \ell \pi$. The boundary term can not be neglected because it contributes to the boundary dynamics.
For the Euclidean signature the gauge group of interest is $SO(3)$, therefore, the Killing-Cartan
metric can be normalized to be $k_{IJ}=\delta_{IJ}$.
Moreover, the connections $A_+$ and $A_-$ as functions of the auxiliar fields, $X^I$ and $Y^I$
respectively, have the same form given in (\ref{solution}). Therefore,
the boundary field theory induced by (\ref{cs-g}) is essentially two theories
like (\ref{nlsm+}) plus one interaction term. This system is currently being analyzed
in \cite{gallardo}.

To conclude, we would like to mention some advantages and disadvantages of our method
by confronting it with some previous works. The procedure described in section 4
has the following advantages: 1) it is covariant because we use the solutions of the equations
of motion in contrast to what happens in Refs. \cite{moore,eli}, where the solution of
the constraints is used, which breaks down covariance. Therefore, in the our approach,
the implementation of other interactions is straightforward while in the later is not. 2) The
resulting boundary action principle is composed entirely by a surface term from the
very beginning in opposition to the resulting action principle of \cite{moore,eli} that
is composed by one surface term and one bulk term, and additional handling is needed to write the
bulk term as a surface term in order to have a boundary field theory. 3) The
implementation of the boundary conditions is systematic because we add them via lagrange
multipliers or as primary constraints in the Hamiltonian formalism. Thus, Dirac method can be
systematic applied to analyze the constraint structure of the theory. In opposition,
in \cite{moore,eli} it is not clear how to incorporate the boundary conditions. 4) we do not have
a criterium to know {\it a priori} to which type of Lagrangians the method described in section 4 can
be applied, besides the Chern-Simons theory. In this sense this a disadvantage of our approach.

\section*{Acknowledgements}
This work was presented in a poster session of the 19th International Conference
on General Relativity (GR19) held in Mexico City, 2010. We thank M\'aximo
Ba\~nados, Jos\'e D. Vergara, and Alejandro Perez for very useful comments. This
work was supported in part by CONACYT, Mexico, Grant No. 56159-F.


\end{document}